\documentclass[%
reprint,
superscriptaddress,
showpacs,
 amsmath,amssymb,
 aps,
pra,
]{revtex4-1}

\usepackage{graphicx}
\usepackage{dcolumn}
\usepackage{bm}

\begin{document}


\title{A phase operator for photons}

\author{Chandra Prajapati}
\affiliation{
 Department of Physics, Indian Institute of Technology Delhi, Hauz Khas, 110016, New Delhi, India}%
\author{D. Ranganathan}%
 \email{dilip@physics.iitd.ac.in}
\affiliation{
 Department of Physics, Indian Institute of Technology Delhi, Hauz Khas, 110016, New Delhi, India}%

\date{\today}

\begin{abstract}
We define a unitary phase operator for photons in a single momentum mode. The operator acts on a  Hilbert space with basis consisting of all number states in both polarizations. The Susskind Glogower operator, ${\hat{E}} = e^{i \hat{\phi}}$ , thus defined is unitary and acts as a true ladder operator over all the states. It satisfies the commutation relation $\left[ {\hat{E}} , {\hat{n}} \right] = {\hat{E}}$; consequently, the phase $\hat{\phi}$ satisfies the canonical commutation relation $\left[{\hat{n}}, {\hat{\phi}} \right] = i$. The eigenstates of the Susskind Glogower operator are found.  Our model is able to directly account for phase measurements when polarization changes cannot be neglected and at the same time gives a well defined phase operator.
\end{abstract}

\maketitle


\section{\label{sec:introduction}Introduction}

The idea of a phase operator for oscillators and fields has a long history. Dirac \cite{dirac}, in an attempt to define
the quantum equivalents of action and angle variables, decomposed the annihilation operator for the harmonic oscillator as $\hat{a} = \hat{E}\sqrt{\hat{n}}={e^{i\hat{\phi}}}\sqrt{\hat{n}}$. Here $\hat{\phi}$ is the phase operator and $\hat{n}$ the number operator of the oscillator. The number operator was to be the quantum equivalent of the classical action and satisfy canonical commutation relations with the phase, $\left[{\hat{n}}, {\hat{\phi}} \right] = i$. As he pointed out, this phase operator is not well defined as there is a lower bound on the eigenvalues of the number operator. In addition, the phase is periodic and defined modulo $2\pi$. \\

As almost all measurements on electromagnetic waves are measurements of intensity or phase; there has been an abiding interest in a phase operator for light too. A large number of subsequent suggestions have been made for the quantum phase operator. Susskind and Glogower \cite{susskind} suggested using the corresponding exponential operator $\hat{E}$ as a ladder operator, $\hat{E}|n>=|n-1>$.  As the operator $\hat{E}$ annihilates the vacuum state to give 0; it is not a unitary operator.  Carruthers and Nieto \cite{carruthers} suggested using the sine and cosine operators, which could be formed out of the exponential operator and its hermitian adjoint but pointed out that they do not commute. Carruthers and Nieto also pointed out that the quantum number specifying the states was closely connected with the spin angular momentum component of the beam along the direction of propagation, so extending the energy spectrum to negative values would solve the unitarity problem.  However, they rejected this extension as negative energies are not possible. This concept of doubling the Hilbert space was revived by Newton \cite{newton}, who also rejected the negative energy states but suggested a formal renumbering of the energy state with the states containing $n$ quanta of negative energy being labeled $-(n+1)$. Modifications of this procedure were suggested by others \cite{garrison, barnett, luis, popov}. These approaches suffer from either lack of unitarity of the exponential operator or failure to obtain the canonical commutation relations for the number operator with the phase operator. An excellent review of the above ideas as well as those to be described below was given by Lynch \cite{lynch}.\\

 Pegg and Barnett \cite{barnett, pegg} suggested the most popular way round these difficulties. The idea is to truncate the oscillator states to a finite number say $s$. The operators in the resulting finite dimensional Hilbert space are well behaved, this is extremely useful as a practical tool. However, questions have been raised \cite{luis, bergou, vourdas} of whether such a weak convergence to the limiting infinite dimensional operator gives the same results as the calculations directly using infinite dimensional operators. In view of these difficulties, Noh,  Fougeres and Mandel \cite{noh, noh1} even suggested that a unique phase operator should not be sought. They suggest that we should only use an operational definition of phase based on experiment performed.\\

With the recent development of quantum tomographic methods to study entanglement and deduce the Wigner distribution, the phase or more accurately the phase difference has again become very important \cite{braunstein}. There has been many subsequent efforts to define a phase operator both for photons and for the closely related problem of the Harmonic oscillator \cite{ban, daoud, xian, hradil, garcia, shunlong, bjork}. Many of these, especially the work by Ban, Luis and Hradil, use two mode operators which are especially well suited to study parametric processes and higher order interference but are also used to cover the case of two orthogonal light polarization modes. However as pointed out by Bjork et. al., these too are not well defined in the vacuum state and indeed require a separate definition of the two mode vacuum state. There is also a related difficulty, the two boson representations used are isomorphic to the angular momentum group, the Cartesian components of which do not commute. It is known that the the Cartesian components of the light polarization need to commute to give a definition consistent with Maxwell's equations \cite{li, vanenk}. Note that these quanta have been measured and shown to add algebraically, with quanta of azimuthal angular momentum \cite{allen}.\\

Sperling and Vogel \cite{sperling} recently pointed out that removal of a right circularly polarized photon is the same as the addition of a left circularly polarized photon. They then suggest treating the left circularly polarized photons as antiparticles of the right circularly polarized photons and used this to construct a unitary Susskind Glogower operator $\hat{E}$. The difficulty with this theory is that there is a difference in angular momentum between the vacuum for left circularly polarized photons and that for right circularly polarized photons.

We suggest a modification of the Sperling-Vogel procedure, based on two requirements. First, the quantum phase being an angle,\textit{ the corresponding momentum must be an action variable}.  The number operator acting on a Fock state then measures the number of action quanta in the state. No experiment directly measures the action, but the number of action quanta is equal to the number of azimuthal angular momenta in that state, which is measurable so we have a suitable proxy. The essence of our suggestion is that action and phase form a conjugate pair and energy and time form another conjugate pair and treating them as freely exchangeable may be part of the reason for our difficulties.

Second, by passage through a phase plate, it is possible to transform a state of definite momentum and polarization ``a mode'', into one of the orthogonal polarization, ``an orthogonal mode'', while preserving the phase or changing it trivially. \textit{ So  the Hilbert space relevant to phase measurements is that spanned by both the polarization modes taken together.} 

The Hilbert space we consider is a tensor product of a two internal dimensional space of polarization and the Fock space of a single polarization mode. This agrees with the representations found by Wigner, \cite{bargmann, wigner} for massless relativistic particles. As Bargmann and Wigner showed the appropriate relativistic representation for photons is two dimensional, this was  elaborated later \cite{kim} to show that it is indeed the natural representation to take into account all polarization phenomena. As the detailed calculations of Kim et. al. show the operators for the commonly used devices like polarizers, phase plates are particularly simple in this representation being constant matrices. The action of such operators is similar on states with the same polarization but different number of photons. Such state independent, passive operators are far more difficult to construct in the two mode theories of polarization.\\

This method permits a natural extension to cover the negative integers. If the number of right circular polarized quanta present is denoted by $ n_{+}  \geq 0$ then the number of left circular quanta present is obviously $ n_{-} \leq 0 $ or vice versa. Newton \cite{newton}, first suggested the use of such a two dimensional representation, but in terms of energy eigenstates, consequently had to suppress the negative energy states. Our interpretation though mathematically very similar is in terms of polarizations, where the internal two space is essential. Thus our model is able to directly account for phase measurements when polarization cannot be neglected while also giving a well defined phase operator.

In the next two sections, the Hilbert space and the phase operator are defined  and the operator is shown to be Hermitian while the corresponding exponential Susskind Glogower operator is unitary. We then find the eigen states of the Susskind Glogower operator and show they are orthogonal. All the previous results cited above can be recovered by restricting our space to states of a single polarization. While we outline our results using the two opposite circular polarizations, they hold in any polarization basis.

\section{\label{sec:state-space}The State Space}

Consider an electromagnetic field propagating in a vacuum, the  states are completely specified by the momentum and the spin angular momentum component (or polarization) along  the direction of propagation \cite{bargmann, mandel}.  The positive frequency components of a single momentum mode of the radiation field are usually written as,
\begin{equation}
\label{field}
\hat{E}(\vec{r},t) \propto \hat{a}_{+} e^{i({\vec{k}\cdot\vec{r}-\omega t})}+{\hat{a}_{-}} e^{i({\vec{k}\cdot\vec{r}-\omega t})}.
\end{equation}

Here $\hat{a}_+$ is the annihilation operator for right circularly polarized light, and ${{\hat{a}}}_-$ is the annihilation operator for left
circularly polarized light.  These are subject to the commutation relations,
\begin{eqnarray}
\left[\hat{a}_{+},{{\hat{a}}^{\dagger}}_{+}\right] =\left[\hat{a}_{-},{{\hat{a}}^{\dagger}}_{-}\right]=1, \nonumber \\
\left[\hat{a}_{+},{\hat{a}}_{-}\right] =\left[\hat{a}_{+},{{\hat{a}}^{\dagger}}_{-}\right]=0.
\end{eqnarray}

There is a modified representation, valid for zero mass particles, found by Wigner \cite{wigner,kim}, which we give below. The two opposite circularly polarized beams are distinguished by their helicity. As there are two states for a given momentum $k$, we write these as a single state with an internal degree of freedom.
\begin{equation}
\left[\begin{array}{c}
 |n_{+}> \\ |n_{-}> 
 \end{array} \right].
\label{state-space}
\end{equation}

They are eigenstates of the Pauli Lyubanski helicity operator$ \hat{P}_{PL}$,
\begin{eqnarray}
\label{counter}
\hat{P}_{PL}\left[\begin{array}{c} |n_{+}> \\ 0 \end{array}\right] = +\left[\begin{array}{c} |n_{+}> \\ 0 \end{array}\right],\nonumber \\
\hat{P}_{PL}\left[\begin{array}{c} 0\\ |n_{-}>  \end{array}\right] = -\left[\begin{array}{c} 0\\ |n_{-}>  \end{array}\right].
\end{eqnarray}
 
 \section{\label{sec:phase-operator}The Phase Operator}

To incorporate the polarization directly into our description, we note that the commutation relations for the annihilation and creation operators imply that if $|n>$ is an eigenstate of the number operator with eigenvalue $n$ then
\begin{eqnarray}
\label{qnumbers}
{\hat{a}}^{\dagger} \hat{a} (\hat{a}|n>) &=& (n-1)(\hat{a}|n>),\nonumber\\ 
{\hat{a}}^{\dagger} \hat{a} ({\hat{a}}^{\dagger}|n>) &=& (n+1)({\hat{a}}^{\dagger}|n>). 
\end{eqnarray}
Both are eigenstates with eigenvalues $n-1$ and $n+1$. \textit{The commutation relations impose no further restrictions on $n$}. The series termination requirement yields the fact that $n$ must be an integer. However the set of all nonnegative integers as well as the set of all non positive integers both satisfy equation (\ref{qnumbers}).
\begin{eqnarray}
{\hat{a}}_{+} |n_{+}>& = &\sqrt{n_{+}}\,|n_{+} -1>,\nonumber \\
{{\hat{a}}_{+}}^{\dagger} |n_{+} > &= &\sqrt{n_{+} +1}\,|n_{+} +1>,\nonumber \\
{\hat{a}}_{+} |0> = 0,\;\; &\Rightarrow& n_+ = 0, +1,+2,+3 \dots, \\
{\hat{a}}_{-} |n_{-}> &=& \sqrt{|n_{-}|}\,|n_{-} -1>,\nonumber \\
{{\hat{a}}_{-}}^{\dagger} |n_{-}> &=& \sqrt{|n_{-}+1|}\,|n_{-}+1>, \nonumber \\
{{\hat{a}}^{\dagger}}_{-} |-1> = 0,\;\; &\Rightarrow& n_{-} = -1,-2,-3 \dots .
\end{eqnarray}
In the subspace $n_{+} \geq 0$, the ${\hat{a}}_{+}$ acts as a annihilation operator, while in the subspace $n_{-} < 0$ the annihilation operator is ${{\hat{a}}^{\dagger}}_{-}$. The choice of quantum number sequence for the negative quantum numbers is the same as that used by Newton, \cite{newton, sperling} and serves to emphasize the similarity to the familiar case of positive quantum number states. It should be emphasized that the state $|-1>$ is the vacuum state for the lower polarization. 
Thus the state space is spanned by the basis,
\begin{equation}
\begin{array}{ccc}
\left[\begin{array}{c}
 |n> \\ 0 
 \end{array} \right],\; n \geq 0,\; \mathrm{and}\;\;
  \left[\begin{array}{c}
 0 \\ |n> 
 \end{array} \right],\; \; n < 0.
  \end{array}
\label{basis}
\end{equation}
Note that though circular polarizations are the natural basis of the helicity operator, the above representation can be used with any other polarization basis.

Any of the operators in a definite polarization subspace can now be extended to the doubled space of both polarizations  as,
\begin{equation}
\label{modified}
     \left[ \hat{a}_{+}, \hat{P}_{PL} \right]=0,  \;\; \left[ \hat{a}_{-}, \hat{P}_{PL} \right]=0. 
\end{equation}

 For any given $n$, the state $n = n_{+} $ and $n = |n_{-}+1|$ contain the same number of photons but have opposite polarizations. It is important to note that both ${\hat{a}}_{+}$ and ${{\hat{a}}^{\dagger}}_{-}$ are operators which reduce the number of photons in their subspace.  That is their repeated application will cause all states with a finite number of photons to be annihilated.\\ 
 
  The method can be used to extend the operation of all operators in either subspace to cover the whole of the doubled space. So far however, the extension is purely formal as we have no operator which connects states labeled by $n_{+}$ and those labeled by $n_{-}$. Consider a new operator,
 \begin{eqnarray}
 \label{flipper}
 {\hat{a}}_{v}=\left(\frac{I-{\hat{P}}_{PL}}{2}\right) &|-1><0|& \left(\frac{I + {\hat{P}}_{PL}}{2}\right),\nonumber\\
 ({\hat{a}^{\dagger}}_{v}){{\hat{a}}_{v}} = |0><0|,& \;&  ({\hat{a}^{\dagger}}_{v})^2 =0 , \nonumber \\
 {{\hat{a}}_{v}}({{\hat{a}^{\dagger}}}_{v}) = |-1><-1|, &\;\;& ({\hat{a}}_{v})^2 =0. 
 \end{eqnarray}
 
This operator causes the vacuum of the right circular polarized states to become the vacuum of the left circular polarized states. A physical realization of this is the passage of a polarized vacuum state through a half wave plate. The hermitian adjoint couples the subspaces in the reverse direction.  We use equations (\ref{modified}) and (\ref{flipper}) to define a lowering operator  which acts on every state in the doubled Hilbert space $H$,
 \begin{equation}
 \label{lowerer}
\hat{a}_m = \left[ \begin{array}{cc}
 \hat{a}_{+} & 0\\ 
  \hat{a}_v  & {{\hat{a}}}_{-}
 \end{array} \right].
\end{equation}
 The commutation relations for the modified annihilation and creation operators,
 \begin{equation}
 {\label{commutator}}
  \left[{\hat{a}}_{m},{{\hat{a}}^{\dagger}}_{m}\right] =
  \left[ \begin{array}{cc}
    {\hat{P}}_{PL} -|0><0| & \;\;\; 0 \\
 0\;\;\;\;& {\hat{P}}_{PL} + |-1><-1|
 \end{array} \right]. 
 \end{equation}
 It is instructive to find the matrix elements of equation (\ref{commutator}) between all the states. Obviously, the non-diagonal elements all vanish, while the diagonal elements yield
 
\begin{eqnarray}
\label{onevacuum}
 <n_{+}|\left[{\hat{a}}_{m},{{\hat{a}}^{\dagger}}_{m}\right]|n_{+}> &= & 1,\nonumber \\
 <n_{-}|\left[{\hat{a}}_{m},{{\hat{a}}^{\dagger}}_{m}\right]|n_{-}> &= & -1,\nonumber \\
 <-1|\left[{\hat{a}}_{m},{{\hat{a}}^{\dagger}}_{m}\right]|-1> = &<0|&\left[{\hat{a}}_{m},{{\hat{a}}^{\dagger}}_{m}\right]|0>=0. 
 \end{eqnarray}

 We see that the usual commutation relations between annihilation and creation operators hold, with the very interesting exception that the annihilation and creation operators commute with each other between the vacuum states $|0>$ and $|-1>$. For a harmonic oscillator, it is the commutation relations eq. (\ref{qnumbers}), which give rise to the energy difference between the states, so the commutation relation (\ref{onevacuum}) guarantees that there is no difference in energy between the two vacuum states. As the particles are bosons, the vacuum state is then
 \begin{equation}
\frac{1}{2}\left[ \left[ \begin{array}{c}
 |0> \\ 0 \end{array}\right]
 +  \left[ \begin{array}{c}
 0 \\ |-1>  \end{array}
 \right]\right] .
 \end{equation}
 
 Following Newton we write the number operator as,
 \begin{equation}
 \label{number}
  {\hat{n}}_{m}=
   \left[ \begin{array}{cc}
    {{\hat{a}}_{+}}^{\dagger}{{\hat{a}}_{+}}  & 0\\
   0 & {{\hat{a}}_{-}}{{\hat{a}}_{-}}^{\dagger} -1  
   \end{array}\right].
 \end{equation}
 This choice is governed by the fact that our operator must count the number of photons or the \textit{energy}.\\
 
 Next we use equation (\ref{lowerer}) to define a Susskind Glogower operator,
 \begin{equation}
 \label{susskind}
 {\hat{E}}_{m} = \left[\begin{array}{cc}
 {\hat{a}}_{+}\frac{1}{\sqrt{{\hat{n}}_{+}}} & 0\\
 \hat{a}_{v} & {\hat{a}}_{-}\frac{1}{\sqrt{{\hat{n}}_{-}}}
 \end{array}\right].
  \end{equation}
 
 This operator has all the required properties,
 \begin{eqnarray}
 {\hat{E}}_{m}\left[\begin{array}{c} |n_{+}> \\ 0 \end{array} \right] &=& \left[\begin{array}{c}|n_{+} -1> \\ 0 \end{array} \right], \nonumber\\
 {\hat{E}}_{m}\left[\begin{array}{c} 0 \\ |n_{-}>  \end{array} \right] &=& \left[\begin{array}{c} 0 \\|n_{-} -1> \end{array} \right], \nonumber  \label{ladder}\\
 \left[{\hat{E}}_{m}, {\hat{n}}_{m} \right] &=& {\hat{E}}_{m}. \label{dirac-exp}
 \end{eqnarray}
 Along with
 \begin{equation}
 {\hat{E}}_{m}{{\hat{E}}^{\dagger}}_{m} = {{\hat{E}}^{\dagger}}_{m}{\hat{E}}_{m} = 1.
 \label{unitarity}.
  \end{equation}
 
 As the Susskind Glogower operator is unitary from equation (\ref{unitarity}), we can write ${\hat{E}}_{m} = e^{i{\hat{\phi}}_m}$ and see that the hermitian phase operator ${\hat{\phi}}_m$ satisfies the canonical commutation relations
 \begin{equation}
 \left[{\hat{n}}_{m}, {\hat{\phi}}_m\right] = i.
 \end{equation}
 Consequently, we have the canonical representation ${\hat{n}}_{m} = i\frac{\partial}{\partial \phi}$ for the number operator in a basis of $\phi$ eigenstates.
 
\section{\label{sec:The eigenstates}The Eigenstates}
 
  We note that the equation (\ref{dirac-exp}) becomes a differential equation in the canonical representation. As the Susskind Glogower operator is unitary, we can see that all eigenstates of this operator must be periodic with period $2\pi$. Then the Floquet theory of periodic differential equations guarantees that every solution of equation (\ref{dirac-exp}) must be of the form, $e^{i{\phi}_0}e^{i2n \phi}$, where ${\phi}_0$ is arbitrary and $n$ is an integer. As is well known, these functions form a  complete basis set for $\phi$ in the interval $0 + {\phi}_0$ to $2 \pi + {\phi}_0$. So the eigenvalue equation for ${\hat{E}}_{m}$ is
 \begin{equation}
 {\hat{E}}_{m}|\phi> = e^{i\phi}|\phi>.
 \end{equation}
 with the solution
 \begin{eqnarray}
 |\phi>= \frac{1}{\sqrt{2\pi}}\sum\limits^{\infty}_{n = 0}
  \left[\begin{array}{c}
 e^{i(n+\frac{1}{2})\phi} \\ 0 
 \end{array}\right] |n> \nonumber \\
 +\sum\limits^{n = -1}_{-\infty}\left[\begin{array}{c}
 0 \\ e^{i(n+\frac{1}{2})\phi} 
 \end{array} \right]|n>.
 \end{eqnarray}
 
  It is interesting to note that the two vacuum state contributions differ by a phase.
 \begin{equation}
 |{\phi}_{0}> = \frac{e^{i\frac{\phi}{2}}}{\sqrt{2 \pi} }\left[ \begin{array}{c} |0> \\0 \end{array} \right] ,\;\;
 |{\phi}_{-1}> = \frac{e^{-i\frac{\phi}{2}}}{\sqrt{2 \pi} }\left[ \begin{array}{c} 0 \\|-1>\end{array} \right].
 \end{equation}
  This suggests that we may be able to distinguish between them using the polarization shifts at reflection. Further, in agreement with the requirements from energy quantization, the action in the vacuum states is also $\frac{1}{2}$.
  
 By restricting our space to states of a single polarization, we get back the phase operators defined by other authors.  We outlined our results using the two opposite circular polarizations, but they hold in any polarization basis. Detailed calculations including the effects of polarization separation by Fedosov Imbert or Goos Hanchen shifts at reflection from dielectrics will be presented elsewhere.
  
  CP wishes to acknowledge support from the U.G.C. of India.

\end{document}